\documentclass[universe,review,accept,oneauthor,pdftex]{Definitions/mdpi}

\firstpage{1} 
\pubvolume{8}
\issuenum{5}
\articlenumber{257}
\pubyear{2022}
\copyrightyear{2022}
\externaleditor{{Academic Editor}: Firstname Lastname} 
\datereceived{27 January 2022} 
\dateaccepted{20 April 2022} 
\datepublished{21 April 2022} 
\hreflink{https://doi.org/universe8050257} 

\usepackage[english]{babel}

\usepackage{xcolor}

\usepackage[T1]{fontenc}

\usepackage{graphicx}	

\usepackage{dcolumn}
\usepackage{bm}

\usepackage{graphicx}
\usepackage{dcolumn}
\usepackage{bm}

\def\ba{\begin{array}}
\def\ea{\end{array}} 
\def\bea{\begin{eqnarray}}
\def\eea{\end{eqnarray}}
\def\beq{\begin{equation}}
\def\eeq{\end{equation}}
\def\ben{\begin{enumerate}}
\def\een{\end{enumerate}}

\def\brr{\begin{array}}
\def\err{\end{array}}

\def\calMa{{V_{4}}}
\def\chiC{\chi_\S}
\def\rL{r_\Lambda}
\def\chiSW{\chi_{*}}

\def\LCDM{$\Lambda$CDM }
\def\dA{{d\Omega}}

\def\in{{\bf{-}}}
\def\out{{\bf{+}}}

\Title{{How the Big Bang} Ends up Inside a Black Hole}

\TitleCitation{How the Big Bang Ends up Inside a Black Hole}
 
\Author{{{Enrique Gaztanaga}} 
 $^{1,2}$\orcidA{}}

\AuthorNames{Enrique Gaztanaga}

\AuthorCitation{Gaztanaga, E.}

\address{%
$^{1}$ \quad Institute of Space Sciences (ICE, CSIC), 
08193 Barcelona, Spain; {{gaztanaga@darkcosmos.com}} 
 \\
$^{2}$ \quad Institut d Estudis Espacials de Catalunya (IEEC), 08034 Barcelona,  Spain}

\abstract{
The standard model of cosmology assumes that  our Universe began 14 Gyrs (billion years) ago from a singular Big Bang creation. This can explain a vast range of different astrophysical data from a handful of free cosmological parameters. However, we have no direct evidence or fundamental understanding of some key assumptions: Inflation, Dark Matter and Dark Energy.  Here we review the idea that cosmic expansion originates instead from gravitational collapse and bounce. 
The collapse generates a Black Hole (BH) of mass $ M \simeq 5 \times 10^{22} M_{\odot}$ that formed 25~Gyrs ago. As there is no pressure support, the cold collapse can continue inside in free fall until it reaches atomic nuclear saturation (GeV), when is halted by Quantum Mechanics, as two particles cannot occupy the same quantum state.
The collapse then bounces like a core-collapse supernovae, producing the Big Bang expansion. Cosmic acceleration results from the BH event horizon. During collapse, perturbations exit the horizon to re-enter during expansion, giving rise to the observed universe without the need for Inflation or Dark Energy.
Using Ockham´s razor, this makes the BH Universe (BHU) model more compelling than the standard singular  Big Bang creation.}

\keyword{
cosmology; dark energy; general relativity; black holes}

\begin{document}

\section{Introduction}
\label{S:1}


A cosmological model predicts the evolution of the observed Universe given some initial conditions.
The standard cosmological model~\cite{Dodelson,Weinberg2008},  called $\Lambda$ Cold Dark Matter ($\Lambda$CDM), assumes that our Universe expansion began in a hot Big Bang creation at the very beginning of space-time. The \LCDM model explains the formation, composition and evolution of our Universe starting from a quantum fluctuation close to Planck scale. Planck scales are so small that space-time itself has to be treated as a quantum object. Such initial conditions violate energy conservation and are very unlikely as they have a low entropy (~\cite{Dyson,PenroseEntropy}). To address such initial conditions properly, we need a new quantum theory of space-time (Quantum Gravity), which opens the door to brane Cosmology~\cite{2005PhLB..614..125D,10.1143/PTP.105.869}  which is a very exciting idea, but it is hard to test with observations.


Despite these shortfalls, the \LCDM model is very successful. However, this is at the cost of introducing three more exotic ingredients or mathematical tricks: Inflation, Dark Matter and Dark Energy, for which we have no direct evidence or understanding at any fundamental level. 
Are they windows for new discoveries, such as String Theory or new forms of matter/energy, or a signal that the paradigm needs to be replaced? Can we choose some different initial conditions and reproduce the success of the \LCDM model without those exotic fixes  and within the known laws of Physics?


Here, we present a brief review that summarizes several recent papers~\cite{Gaztanaga2020,Gaztanaga2021,FG20,hal-03344159,GaztaUniverse,Benjamin,GF21} that suggest a simpler explanation: the Black Hole Universe (BHU). This review also includes some new results and ideas. Some previous studies misinterpreted super horizon scales as scales that were outside the BHU. This is clarified here together with some new details regarding the Big Bounce and the observational interpretation of super horizon perturbations. In Section~\ref{sec:LCDM} we give a brief presentation of the \LCDM model and its observational support. 
In Section~\ref{sec:BHU} we present the BHU model using a Newtonian approach. In Cosmology, one can use Newtonian physics to model to a  good approximation of both the background and its perturbations ~\cite{Bernardeau}. A consistent Newtonian version of the BHU solution is a good indication that we understand the physics involved. Appendices \ref{app:exactGR} and \ref{app:lambda} present a summary of the same BHU solution in the more rigorous GR approach.
Appendix \ref{app:rot} presents some new considerations of the possible effect of rotation in the BHU solution.
We end with a Summary and Discussion that includes a review of related literature and previous results and a comparison between models.

\section{Observational Evidence for \LCDM} 
\label{sec:LCDM}

We briefly discuss here the main observational evidence of the \LCDM, focusing on why exactly it needs those fixes. This review is not exhaustive and does not include all the relevant references. It is just a brief introduction and further work can be found in the references within. We assume flat topology
(we will explain why in \S4).
 
\subsection{The Expansion of the Universe and the FLRW Metric}

In 1929, Edwin Hubble published~\cite{Hubble1929} his famous diagram or linear relation (the Hubble law): $\dot{r} = H r$ 
relating the radial distance $r$ of 46 galaxies to their radial recession velocity $\dot{r} \simeq z c$, given by the redshift $z$ and the speed of light $c$ ($\dot{r}$ is the time $\tau$ derivative: $\dot{r} \equiv \frac{dr}{d\tau}$). Hubble used redshift $z$ from galaxy spectra estimated and published by Vesto Slipher (1917)~\cite{Slipher1917} and Cephid distances $r$ developed by Henrietta Leavitt~\cite{1912Leavitt} and calibrated by E. Opik~\cite{Opik1922}. However, it was George Lemaitre who first understood, in 1927~\cite{Lemaitre}, the meaning of such a discovery~\cite{Elizalde2021}: that spacetime is expanding following the new theory of General Relativity (GR) by Albert Einstein~\cite{Einstein1916}. 

However, you do not actually need GR to figure out the expansion equations.
At large scales, the observable Universe looks homogeneous and isotropic. This alone tells us
that a physical radial distance $r$ has to scale as  $r=a(\tau)\chi$, where $a(\tau)$ is a dimensionless scale factor and $\chi$ is a comoving coordinate, which is fixed ($\dot{\chi}=0$) for any comoving observer like us, moving with the expansion. This is the  Friedmann–Lemaitre–Robertson–Walker (FLRW) metric (i.e., Equation~(\ref{eq:frw})).
The  observed expansion law follows from derivation: $\dot{r}=\dot{a} \chi = H r$ where $H \equiv \dot{a}/a$ is the Hubble expansion rate. Nowadays, $H$ is measured to be $H_0 \simeq 70$ Km/s/Mpc, so a galaxy at $r\simeq 300$Mpc has a recession velocity of $\dot{r} \simeq z c$ with a redshift $z \simeq 0.07$. The Universe was 7\% smaller at the time the light from that galaxy was emitted, $\tau \simeq  \frac{r}{c} = 92$Myr ago.  The expansion time is $H_0^{-1} \simeq 14$Gyr. 

Consider a spherically symmetric region of space $r<R$ with a fixed mass $M$ (such as Lemaitre model~\cite{Lemaitre}). We can use Gauss law (or the corollary to Birkhoff theorem in GR~\cite{Deser2005}) to ignore what is outside $R$ so the dynamics of $R$ will be given by the free-fall equation:
\beq
E=\Phi +K =0 \, \Rightarrow \, 
K = \frac{1}{2} \dot{R}^2 =   \frac{1}{2} H^2 R^2 = - \Phi = \frac{GM}{R} = \frac{4\pi G}{3} \rho R^2. 
\label{eq:phi}
\eeq
{The} above equation leads to: 
\beq
r_H^{-2} \equiv  H^2(\tau) = \frac{8\pi G}{3} \rho(\tau), 
\label{eq:H2}
\eeq
which is independent of $R$. This simple Newtonian derivation reproduces exactly the full solution to GR field equations  (i.e., Equation~(\ref{eq:Hubble})).
At any time, the expansion rate $H^2$ is given by
$\rho$. In our Universe we have measured their values today ($\rho_0$ and $H_0$) to find that they do follow: 
$H_0^2 \simeq 8\pi G\rho_0/3$.
We use units where the speed of light is $c=1$, and $r_H \equiv H^{-1}$ is called the Hubble Horizon because it corresponds to an expansion velocity $\dot{r} = H r$ equal to the speed of light ($\dot{r}_H = 1$).
Energy--Mass conservation requires
$\rho \propto a^{-3(1+\omega)}$, where $\omega=p/\rho$ is the equation of the state of the fluid:
$\omega=0$ for matter, $\omega=1/3$ for radiation and $\omega=-1$ for vacuum. 
Given $a_*=a(\tau_*)$  at some reference  time $\tau_*$ and
$\tau=0$ at $a=0$, the solution to Equation~(\ref{eq:H2}) for one component is:
\beq
H^2= H^2_* \left(\frac{a}{a_*}\right)^{-3(1+\omega)}  \Rightarrow  a(\tau) = a_* \left[\frac{3(1+\omega)}{2} \tau H_*\right]^{\frac{2}{3(1+\omega)}}
\Rightarrow r_H= \frac{3(1+\omega)}{2}\tau . 
\label{eq:collapse}
\eeq
{During} collapse, $H$ and $\tau$ are negative.
Note that $r_H \propto a^{3(1+\omega)/2}$, so for regular matter ($\omega>0$), it grows faster than comoving scales: $r = a \chi$ (the opposite is true for $\omega=-1$). Thus, for $\omega>0$ all scales become super horizon ($r>r_H$) during collapse ($H<0$) and re-enter the Hubble horizon during expansion. Note that $r_H$ 
increases with time $\tau$ (like the particle horizon). These equations are the exact solutions to GR for an FLRW metric, where $\tau$ is the proper time for a comoving observer. Using Equations~(\ref{eq:H2}) and (\ref{eq:collapse}) we find:
\beq
\rho = \frac{ \left[(1+\omega) \, \tau \right]^{-2}}{6\pi G}
\simeq 1.3 \times 10^{-12} \frac{M_{\odot}}{\text{Km}^3} \,  \left[\frac{(1+\omega) \tau}{\text{seconds}}\right]^{-2},
\label{eq:rho}
\eeq
which tell us what the density is at any time $\tau$. In general, $\rho$ could be made of several components $\rho_i$: $\rho = \sum_i \rho_i$, each with different $\omega_i$. The relative contributions are called cosmological parameters: $\Omega_i \equiv \rho_i/\rho$, so that $\sum_i \Omega_i =1$. As $\tau \Rightarrow 0$, the matter density becomes very high and the radiation density dominates as temperature increases: $T= T_0/a$.

\subsection{Nucleosynthesis and CMB}

In 1964, Penzias and Wilson~\cite{Penzias} accidentally found a uniform Cosmic Microwave Background (CMB) radiation of temperature $T_0 \simeq 3K$. Robert Dicke, James Peebles, P. G. Roll, and D. T. Wilkinson in the companion publication~\cite{Dicke} interpret this radiation as a signature from the hot Big Bang: the oldest light in the Universe.
This was first noticed in 1948 by R.Alpher and R.Herman~\cite{1948PhRv...74.1737A,Alpher} who developed the theory of the primordial nucleosynthesis and predicted a leftover CMB radiation of  $T_0 \simeq 5K$, closed to the observed value. 
The idea behind it is simple. Because the universe is expanding, when you imagine going back in time the density must become higher and higher, atoms will break and the resulting plasma will be dominated by radiation, like the interior of a star. If you simulate an expansion from such initial conditions you can build a prediction for the primordial abundance of elements and radiation that we observed today. This is called primordial nucleosynthesis.

Hydrogen is the most abundant
element measured in our Universe. 
Around $\sim 75$\% of the total mass of the atoms (nucleons) in the Universe is in the form of  hydrogen, the remaining $25$\% is mostly Helium. The abundance predicted by nucleosynthesis
depends on the cross section of several  Nuclear Physics reactions, such as neutron capture or decay. These are proportional to the ratio $\eta = \rho_B/\rho_R$ of the number density of baryons $\rho_B$ (protons and neutrons) to that of photons, $\rho_R$, given by the CMB background temperature $T=T_0/a$. So a measurement of the primordial element abundance and $\rho_B$ can be used to predict $T_0$. Nowadays, we use 
the more precise measured value $T_0=2.726K$ and the observed abundance to predict $\rho_B$.
In relative units: $\Omega_B=\rho_B/\rho \simeq 0.05$~\cite{Steigman}. So that only $\simeq 5\%$ of the total energy-density in our Universe is made of regular matter (i.e., made of known baryons and leptons). The rest, according to \LCDM, is made of Dark Matter and Dark Energy. Where do these numbers come from? Why 5\% and not 20\%?

\subsection{Cosmic Inflation and the Horizon Problem}

Cosmic Inflation~\cite{Starobinski1979,Guth1981,Linde1982,Albrecht1982} consists of a period of exponential  expansion that must have happened right after the beginning of time ($\tau=0$). There are over a hundred versions and variations~\cite{Liddle1999,Weinberg2008},  but generically the model requires some hypothetical new scalar field (the inflaton) with 
negligible kinetic energy ($\omega=-1$) to dominate the very early universe. After expanding by a factor of $e^{60}$, inflation leaves the universe empty  and we need a mechanism to stop inflation and create the matter and radiation that we observe today. This is called re-heating. All these components require some fine tuning and free parameters that are hard to test because the physics involved is beyond reach by experiments \cite{Dodelson}. However, Inflation solves some important mysteries that we do not know how to fix otherwise within \LCDM.

As mentioned below Equation~(\ref{eq:collapse}),
the structures that we observe today were not in causal contact in the past. We say that they are  super horizon scales. Structures that are larger than $r_H$ cannot evolve because the time a perturbation takes to travel that distance is larger than the expansion time. How can these structures form if they were not in causal contact? This is the horizon problem. 
A clear evidence of this problem is the uniform CMB temperature across the full sky. The Hubble horizon $r_H$ at CMB times only subtends about one degree in our sky. So causality cannot explain the observed all sky CMB uniformity.
The horizon problem is solved by inflation because, during inflation, structures of all scales become a super horizon. After inflation ends, they re-enter the horizon  ($\omega=-1$). Moreover, reheating provides a very uniform temperature background at the end of inflation.

\subsection{Structure Formation and Dark Matter}

In 1992, NASA's Cosmic Background Explorer (COBE) satellite detected temperature variations of very small relative amplitude $\delta_T \simeq 10^{-5}$ in the CMB~\cite{COBE}.
We believe those were the seeds that grew under gravitational collapse 
from to form stars, galaxies and the cosmic web that we observe today. However, where do the seeds come from? Models of Inflation propose that these seeds come from super horizon quantum fluctuations that were exponentially inflated during inflation. Inflation predicts a power law (almost scale invariant) spectrum of fluctuations  which agrees with the shape measured by later CMB missions~\cite{WMAP,P18cosmo,ACT,SPT2018} and clustering in Galaxy Surveys~\cite{1990Natur.348..705E,GB1998,Gawiser,DES2021}. However, inflation does not provide a specific prediction for $\delta_T \simeq 10^{-5}$: it is just a free parameter of the model.

The measured $\delta_T \simeq 10^{-5}$  is too small to explain the observed structure in galaxy surveys today~\cite{1992MNRAS.258P...1E,1998ApJ...499..526T}. Some fix is needed: galaxy bias~\cite{Fry1993,Juszkiewicz} or a $\Lambda$ term~\cite{1990Natur.348..705E}.
The shape of the spectrum of fluctuations (including the Baryon Acoustic Oscillations, BAO~\cite{EisensteinHu,Eisenstein2005,GCH2009,Gaztanaga2009}) in the CMB  and Galaxy Surveys, also required another free component to agree with the \LCDM model. They require a new type of matter, that we called Cold Dark Matter (CDM~\cite{1985ApJ...292..371D}), which is not made of regular matter (baryons) and interacts very weakly with matter or radiation (thus the name). CDM needs to be about four times more abundant than regular matter: $\Omega_{CDM} \simeq 4 \Omega_B$. Such CDM is also needed to understand the motion of galaxies in clusters~\cite{Zwicky}, the galaxy rotational curves~\cite{Rubin1970}, gravitational lensing~\cite{Bullet}, galaxy evolution~\cite{1985ApJ...292..371D}, cosmic flows~\cite{Feldman} and structure in galaxy maps~\cite{1990Natur.348..705E,GB1998,Gawiser,DES2021}. Despite enormous observational efforts in the last 30yrs, such Dark Matter component has never been directly detected as a real particle or object~\cite{2005PhR...405..279B,2019Univ....5..213P}. 

\subsection{Cosmic Acceleration, Dark Energy and the Static Universe}

Usually, cosmic acceleration is defined by the adimensional coefficient $q \equiv  (\ddot{a}/ a) H^{-2}$. Taking a derivative of Equation~(\ref{eq:collapse}) we find $q= -\frac{1}{2} (1+3\omega)$. For regular matter we have $\omega>0$ so we expect the expansion to decelerate ($q<0$) because of gravity. However, the latest concordant measurements from a Type Ia supernova (SN)~\cite{Perlmutter,Riess}, galaxy clustering and  CMB all agree with an expansion that tends to $\omega = -1.03 \pm 0.03$~\cite{DES2021} or $q \simeq 1$ in our future.

Dark Energy (DE) was introduced~\cite{Huterer} 
to account for $\omega<0$. However, there is no fundamental understanding of what DE is or why $\omega \simeq-1$. A natural candidate for DE is the cosmological constant $\Lambda$~\cite{Einstein1917,Weinberg1989,Carroll1992,Peebles2003}, 
which has $\omega =-1$ and can also be thought of as the ground state of a scalar field (the DE), similar to the Inflaton. $\Lambda$ can also be a fundamental constant in GR, but this has some other complications~\cite{Weinberg1989,Carroll1992,Peebles2003}. Including DE in the \LCDM model is also needed to complete the energy budget for our Universe: 5\% baryons ($\Omega_B \simeq 0.05$), 25\% Dark Matter ($\Omega_{DM} \simeq 0.25$) and 70\% DE ($\Omega_\Lambda \simeq 0.7$), so that $\Omega_B+\Omega_{DM}+\Omega_\Lambda =1$, as needed. DE is also important for understanding the Integrated Sachs--Wolfe (ISW) effect~\cite{Crittenden,Fosalba2003,Seife2003,Gaztanaga2006}, 
and to have a longer age estimate of 14 Gyr, which is needed both to account for 
the oldest stars and to have more time for structures to grow from  the small CMB seeds $\delta_T \simeq 10^{-5}$ to the amplitude (and shape) we observe today in Galaxy Maps~\cite{1990Natur.348..705E, 1992MNRAS.258P...1E}.

Note how $q=1$ means $\dot{H}=0$, so that $H$ becomes constant and all structures become super horizon and freeze, as in Inflation. In the physical or rest frame (see Section~\ref{sec:frame}) this corresponds to a static (deSitter) metric. We are used to repeating that the universe accelerates, but in the limit $q \Rightarrow 1$
it is more physical to say that the universe becomes static, as proposed by Einstein~\cite{Einstein1917} when he introduced $\Lambda$. This can be understood with the Twin paradox analogy of Especial Relativity to explain time dilation. Time happens slower for the comoving observer according to the physical observer at rest. In the limit of exponential expansion, time freezers and the expansion stops. Something that is static for the rest frame observer
happens at constant velocity ($H$ constant) for the comoving observer.

\section{Inside a Black Hole (BH)}
\label{sec:BHU}

In this section we will first present three different arguments
that indicate that our Universe is inside a BH. This will lead to the BH Universe (BHU) model. How did we end up inside a BH? We will conjecture a new start for our Universe that could explain both the Big Bang expansion and why we are inside a BH, without the need to restore to a Quantum Gravity singularity.

\subsection{What Is a BH?}

A BH is an object with a radial escape velocity $\dot{R} = c \equiv 1$.
The  escape velocity $\dot{R}$ is the minimum one needed to just escape the gravitational pull of a mass $M$. 
This requires: $\frac{1}{2} \dot{R}^2  = \frac{GM}{R}$.
Thus, for $\dot{R} = c \equiv 1$ we have that  $R \equiv r_S= 2GM$, which is called the Event Horizon. As events cannot travel faster than $c$, nothing can escape from inside $r_S$.
Thus we define a BH as an object of mass $M$ whose radius is:
\beq
r_s =2GM \simeq 2.9 \text{Km} \frac{M}{M_{\odot}},
\eeq
so that a solar mass BH has a radius of 2.9Km. The density of a BH only depends on $r_S$:
\beq
\rho_{BH} = \frac{M}{V} = \frac{3M}{4\pi r_S^3} =\frac{3r_S^{-2}}{8\pi G}
\simeq 9.8 \times 10^{-3}  \left[\frac{M_{\odot}}{M}\right]^2 \, \frac{M_{\odot}}{\text{Km}^{3}}. 
\label{eq:BHrho}
\eeq
{This} value should be compared to the atomic nuclear saturation density:
\beq
\rho_{NS} \simeq 2 \times 10^{-4} \frac{M_{\odot}}{\text{Km}^{3}}, 
\label{eq:NS}
\eeq
which corresponds to the density of heavy nuclei and results from the Pauli exclusion principle. For a Neutron Star (NS) with $M \simeq 7M_{\odot}$ both densities are the same: $\rho_{BH}=\rho_{NS}$.

\begin{figure}[H]
\includegraphics[width=1.\linewidth]{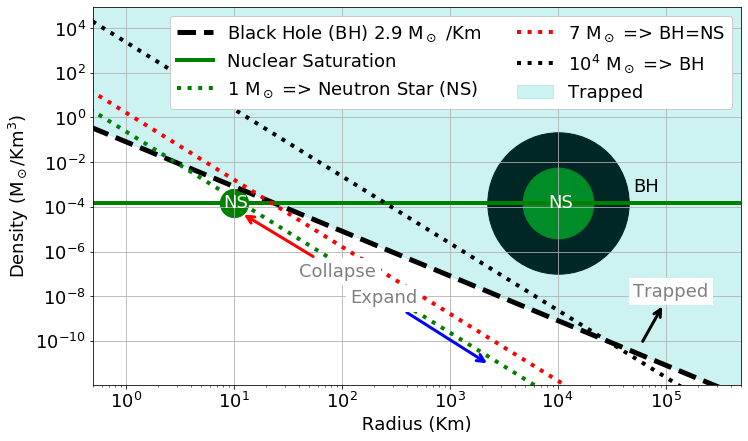}
\caption{Illustration of the collapse of one solar mass (1 $M_{\odot}$, green dotted line) Neutron Star (NS). As the NS collapses the radius $R$ decreases and the density increases as $R^{-3}$. The collapse stops (bounce back or explodes as a supernova) when the density reaches nuclear saturation $\rho_{ns}$ (green horizontal line). For masses larger that 7 $M_{\odot}$ (red dotted line) the cold cloud collapses first into a BH and matter gets trapped inside the event horizon $r_S$ (black dashed line $R^{-2}$). A NS could  collapse inside a larger BH, but it can not escape $r_S$. }
\label{fig:NS-BH}
\end{figure}

This explains why typical NS stars are not larger than $M \simeq 7 M_{\odot}$, as they could collapse first into a BH. This is illustrated by Figure~\ref{fig:NS-BH} which compares the collapse density of a fix mass cold cloud as a function radius to the density of a BH. Because the star is collapsing in freefall (assuming no significant pressure support) nothing prevents the BH to form if the density reaches BH density before it reaches nuclear saturation. 
The maximum observed $M$ for NS is closer to  $M \simeq$ 3 $M_{\odot}$~\cite{2016ARA&A..54..401O} which agrees with more detailed considerations that include the equation of state estimates.
\subsection{Inside a Black Hole}

\begin{figure}[H]
\includegraphics[width=1.\linewidth]{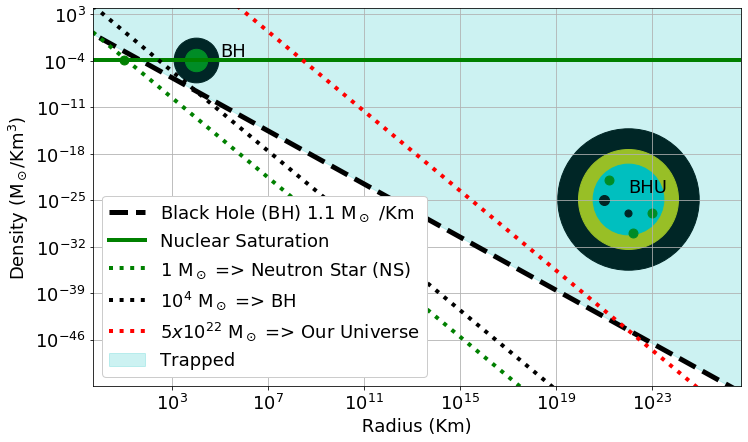}
\caption{This is similar to Figure~\ref{fig:NS-BH} but extending the scale to include a cloud of mass $M = 5 \times 10^{22} M_\odot$ (red dotted line) which corresponds to the size of our Universe.}
\label{fig:NS-BHU}
\end{figure}

The density of a BH in Equation~(\ref{eq:BHrho}) is the exact density of our Universe  in Equation~(\ref{eq:H2}) inside its Hubble Horizon $r_H=1/H$, as for $R=r_H$, the expansion law gives: $\dot{R}= H R = 1$.
So the Hubble volume around us ($R<r_H$) is causally disconnected from the rest ($R>r_H$) and has the density of a BH. 
Very different observations (CMB, SN, BAO, lensing and LSS)
indicate ~\cite{DES2021} that  $H$ tends to a constant $H_\Lambda^2 =   H_0^2 \Omega_\Lambda = \frac{8\pi G}{3} \rho_\Lambda$  (i.e., $\omega=-1$) so the Universe asymptotically  becomes static with a fixed radius ( $r_\Lambda= H_\Lambda^{-1}$). Nothing can escape 
 $r_\Lambda$ and the mass inside is given by:
 \beq
 M = \frac{4\pi}{3} r_\Lambda^3 \rho_\Lambda = \frac{r_\Lambda}{2G}, 
 \eeq
i.e.,:  $r_\Lambda = 2~GM$. This is the definition of a BH. 
So we do live inside a BH of mass and size:
\beq
 M \simeq 5 \times 10^{22} M_{\odot}   \,\,\,;\,\,\, r_S = r_\Lambda = r_H(a=\infty) \simeq 6 \times 10^{22}\text{km},
\label{eq:rs}
\eeq
for $\Omega_\Lambda  \simeq 0.7$ and $H_0 \simeq 70$ Km/s/Mpc. 
Figure~\ref{fig:NS-BHU} compares the BHU formation with that of an NS.
Inside $ r_S \simeq 6 \times 10^{22}
~\text{km}$ the density is very low and nothing can stop further collapse. So NS, galaxies and planets could also eventually form inside a BH.
\subsection{The Black Hole Universe (BHU)}

An FLRW metric with $\Lambda$ 
is inside a trapped surface. The maximum radial distance travel by light (a null geodesic) is:
\beq
r_* = a \int_\tau^\infty \frac{d\tau}{a(\tau)} =
a \int_a^\infty \frac{d\ln{a}}{a H(a)} < \frac{1}{H_\Lambda} \equiv r_{\Lambda}.
\label{eq:chi}
\eeq
{As} time increase, the Hubble rate becomes constant and $r_*$ becomes a constant value $r_* = r_\Lambda$.
No signal from inside $r_*$ can reach outside, similar to in the interior of a BH.

If we use $r_S=2GM$ in Equation~(\ref{eq:phi}) we find:
\beq
R=[r_H^2 r_S]^{1/3}  \Rightarrow R(\tau) = \frac{3(1+\omega)}{2} \,\tau_*^{1/3} \tau^{2/3} = r_S \left(\frac{a}{a_{BH}}\right)^{1+\omega},
\label{eq:R}
\eeq
where $a_{BH}$ is the scale factor when the BH event horizon is reached.
For a regular star $R>r_S$ so the expansion is subluminar  $R<r_H$.
Our Universe has $R>r_H$ (we observe super-horizon scales in the CMB) which, using Equation~(\ref{eq:R}), requires $R<r_S$.
This is a third indication that we are  inside our own BH!

\begin{figure}[H]
\includegraphics[width=.5\linewidth]{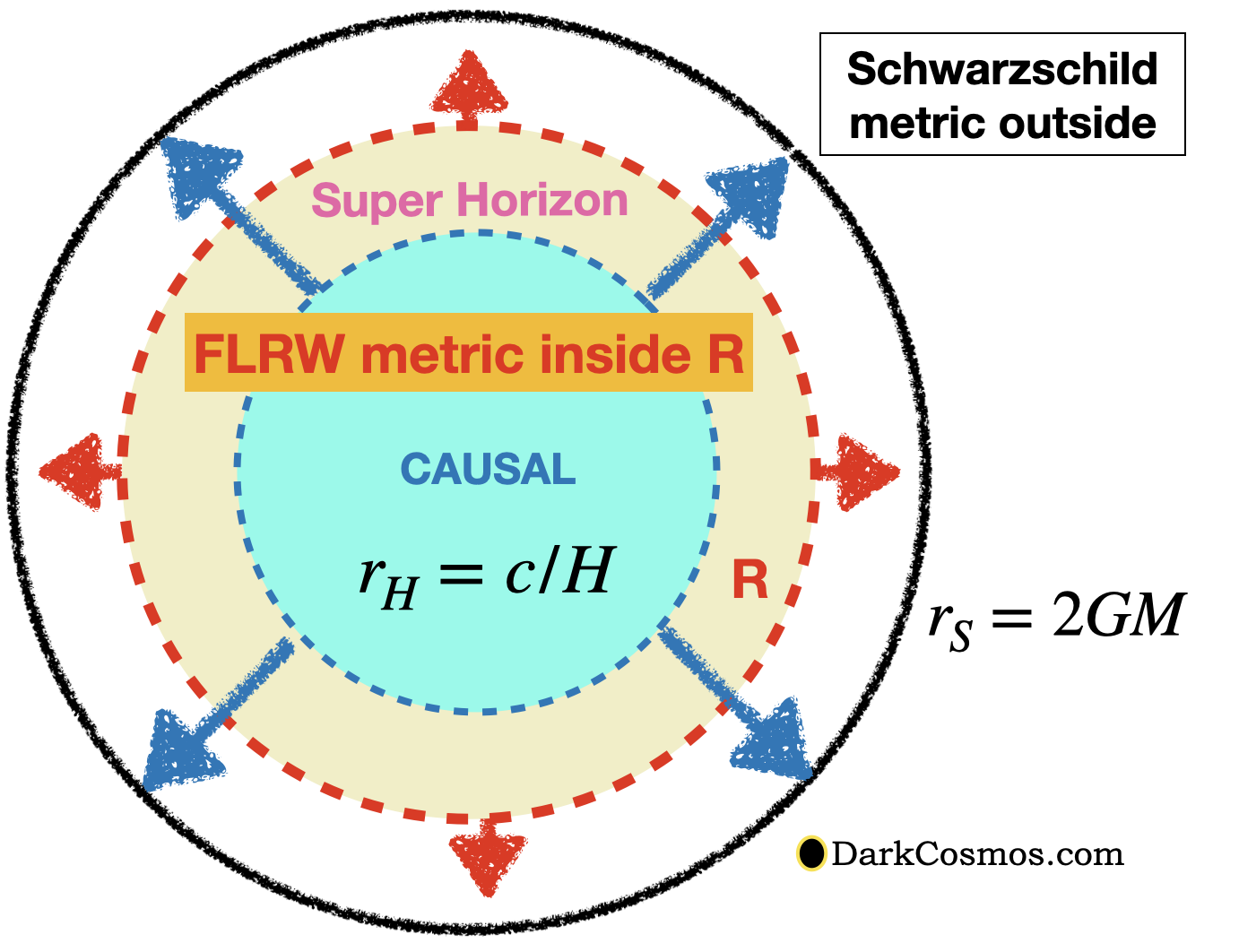}
\caption{Illustration of our Universe inside the event horizon $r_S=2GM$. This is a Schwarzschild (empty) metric outside ($r>R$) and an FLRW metric inside $R$ (red dashed line) with mass $M$. The Hubble radius
$r_H=c/H$ (dashed line) defines the volume inside causal contact (blue shading) from the center. The BHU solution in Equation~(\ref{eq:R}) requires $ R=[r_H^2 r_{S}]^{1/3}$, so that $R$ grows slower than $r_H$.
 There is a region with matter outside the Hubble radius $R>r>r_H$ (yellow shading) with super horizon (or frozen) perturbations. This solves the horizon problem in Cosmology and is a source for perturbations that enter the horizon as the metric expands, creating LSS and BAO in Cosmic Maps,
 pretty much like what is usually assumed for Cosmic Inflation.}
\label{fig:BHgrowth}
\end{figure}

For $\omega=p=0$, $R$ is a time-like geodesic with constant $\chi=R/a=r_S/a_{BH}$. For a null geodesic $R=r_*$ ($\omega \ne 0$) in Equation~(\ref{eq:chi}).
Equation \ref{eq:R} gives the evolution of a finite FLRW cloud radius $R(\tau)$. 
Compared to Equation~(\ref{eq:collapse}) we can see that $R$ grows slower than $r_H$ so perturbations become super-horizon during collapse and re-enter during expansion. So the collapsing phase acts like Inflation.
In units of $r_H$ today $c/H_0 \equiv 1$, at CMB times ($a \simeq 10^{-3}$):  $r_H \simeq 5 \times 10^{-5}$, while $R$ is about 30 times larger.  
For constant  $H=H_\Lambda$ we have $R=r_S$, which is larger than $R_0$ today. 
Note how for $R<r_S$ (i.e., inside the BH)  Equation~(\ref{eq:R})  indicates that there is a region with no matter: $r_S>r>R$ and a region with matter outside the Hubble horizon $R>r>r_H$. 
This is illustrated in Figure~\ref{fig:BHgrowth}.

Here we have obtained Equations~(\ref{eq:H2}) and (\ref{eq:R}) just using Newtonian Mechanics with the definition of a BH. This is the same solution as the BH Universe (BHU)~\cite{hal-03344159}, which is an exact solution to GR and corresponds to an FLRW cloud as in Equation~(\ref{eq:phi}). Appendix \ref{app:exactGR} presents this same BHU solution within GR.

\subsection{How Did We End up Inside a BH?}

\begin{figure}[H]
\centering\includegraphics[width=1.\linewidth]{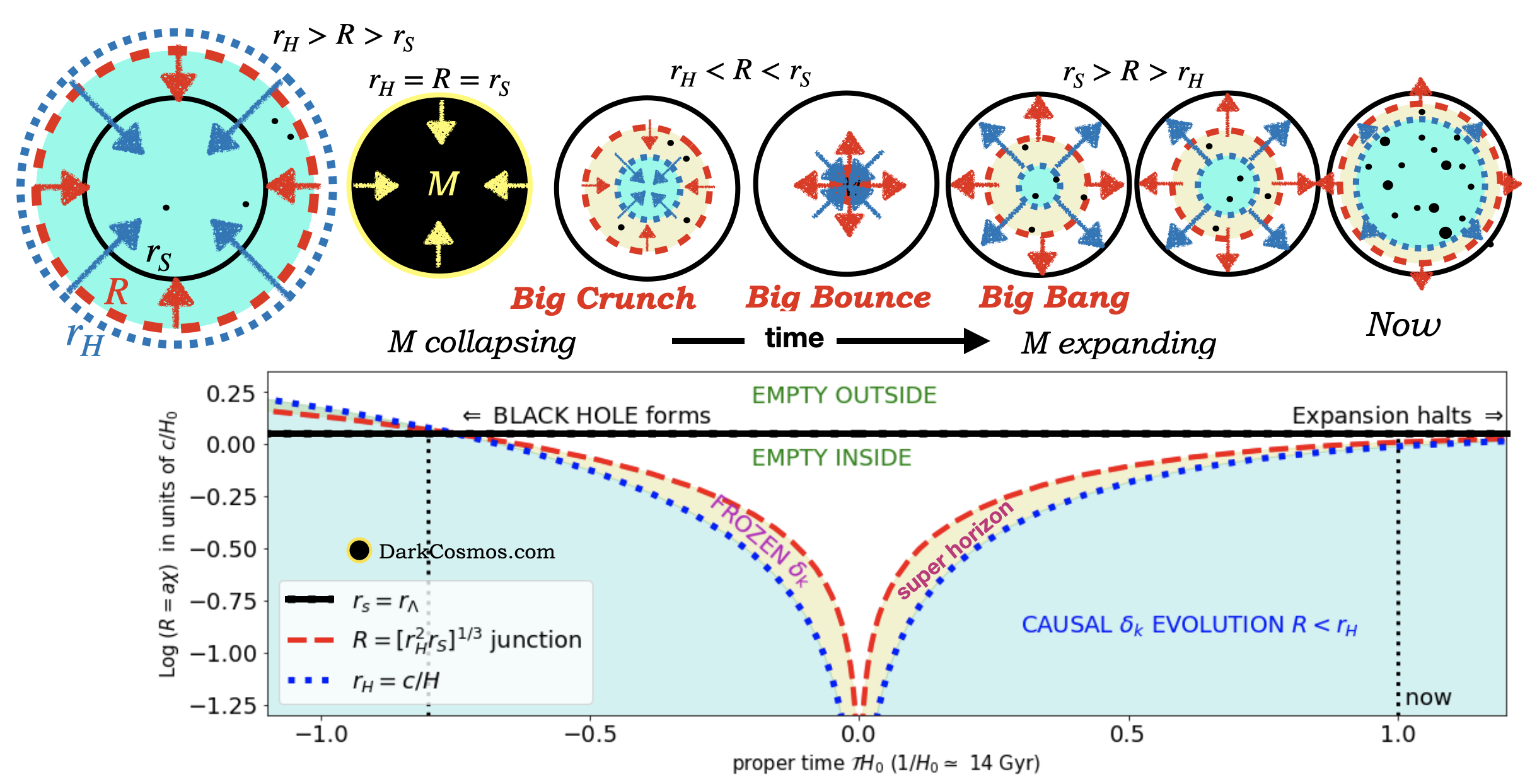}
\caption{Physical coordinate radius $R$ collapsing and expanding as a function of comoving time $\tau$. A spherical cloud of radius $R$ and mass $M$ starts collapsing free-fall under its own gravity. When it reaches $R=r_{S} =2GM$ it becomes a BH (black sphere).  The collapse proceeds inside the BH until it bounces into an expansion (the hot Big Bang). The BH Event Horizon $r_{S}$ behaves like a cosmological constant with $\Lambda= 3/r_{S}^2$ so that the expansion freezes before it reaches back to $r_{S}$.  Blue shading ($R<r_H$) indicates causal evolution of radial perturbations. White is approximated as empty space. Super horizon structures  in-between $R$ and $r_H$  (yellow shading) are "frozen" and they seed structure formation as they re-enter $r_H$.
Contrary to Inflation, the super horizon spectrum of perturbations  in the BHU has a cut-off given by $R$.}
\label{fig:BHevolution}
\end{figure}

Our Universe must have collapsed to form a BH. 
Before it collapsed, the density is so small that there are no interactions other than gravity. Even radiation  escapes the cloud. The density is still very low when $M$ approaches its corresponding event horizon $R=r_S=2GM$, but the gravitational pull is still that of a BH. Radial comoving shells of matter are in free fall collapse, so they do not feel that pull. This is the Equivalence Principle. 
So the collapse continuous pass $R=r_S$ inside the BH.
We take $\tau_*$ in Equation~(\ref{eq:collapse}) to correspond to the time $\tau_{BH}$  when  $r_H = -r_S$, i.e., FLRW cloud formed a BH:
\beq
\tau_{BH} = \tau_* =- \frac{2}{3(1+\omega)} r_S \simeq -11 Gyrs,  
\label{eq:tauBH}
\eeq
where we have used Equation~(\ref{eq:rs}) and $\omega \simeq 0$ (the latest stages of the collapse could have $\omega \simeq 1/3$, but they last a negligible time compare to matter domination).
Thus the BH forms 11 Gyr before $\tau=0$ (the Big Bang) or 25Gyr ago. 
The cold collapse continued after the BH formation.
In the last stages of the collapse atoms could ionized and part of the energy could transform into heat. This could slow down the collapse. Figure~\ref{fig:BHevolution} shows the numerical BHU solution using Equations~(\ref{eq:collapse}) and (\ref{eq:R}).

\subsection{The Big Crunch}

As mentioned before, there is a region outside the Hubble Horizon $R>r>r_H$ which is dynamically frozen (yellow shading in Figures~\ref{fig:BHgrowth} and \ref{fig:BHevolution}). This is the source for super horizon perturbations, which can be observed today in the CMB temperature maps.
Any small irregularities $\delta \equiv \Delta \rho/\rho$ (such as the particle composition of the fluid) will grow under gravity. This is the so-called
gravitational instability. The growth of $\delta$ can start early on within the FLRW cloud, well before $\tau_{BH}$. 
The amplitude of $\delta$ from gravitational instability is  scale invariant ~\cite{Zeldovich1970,Harrison1970,PeeblesYu}. 
In the linear regime, $\delta$  follows a damped harmonic oscillator equation whose solutions 
~\cite{Bernardeau}
are $D_{+} \propto a$ and $D_{-} \propto a^{-3/2}$, which correspond to the growing and decaying mode during expansion.
In the collapsing phase the damping term has a negative sign and fluctuations grow faster with time because $D_{-}$ is the growing mode when $a$ goes to zero.
It is therefore likely that galaxy, stars, planets or life could also form during the collapsing phase. The details might depend on the original FLRW cloud composition. As the cloud collapses and the background density increases, the structures will disappear inside a hot Big Crunch, but the largest scale density perturbations will become super horizon scales (freeze out) and survive the Big Bounce, as they correspond to variations of the background over scales that are causally disconnected.

\subsection{The Big Bounce}

The energy density $\rho$ in Equation~(\ref{eq:rho}) is the same everywhere inside $R$. By $\tau \simeq -10^{-4}$ seconds, $\rho$ approaches nuclear saturation (GeV) in Equation~(\ref{eq:NS}). 
The radius of our Universe  $R$  is close to the distance between Earth and the Sun. However,
the Hubble radius is only few Km (containing a few solar masses). So the physical situation is similar to the interior of a regular collapsing star.  We conjecture that this leads to a Big Bounce because of the Pauli exclusion principle of Quantum Mechanics.
Neutron density is the highest cold density observed in nature.  The collapse is halted by neutron degeneracy pressure, causing the implosion to rebound ~\cite{1979ARA&A..17..415B}.
If the neutron material is elastic enough~\cite{elasticity} the collapse could just bounce into an expansion, pretty much like a bouncing of a ball. However, if the expansion rate is too high, the collapse could also led to a supernova (SN) explosion.
Global rotation of the FLRW cloud, could slow down the expansion rate (see Appendix \ref{app:rot}) and play some role in the bounce.

\begin{figure}[H]
\includegraphics[width=1.\linewidth]{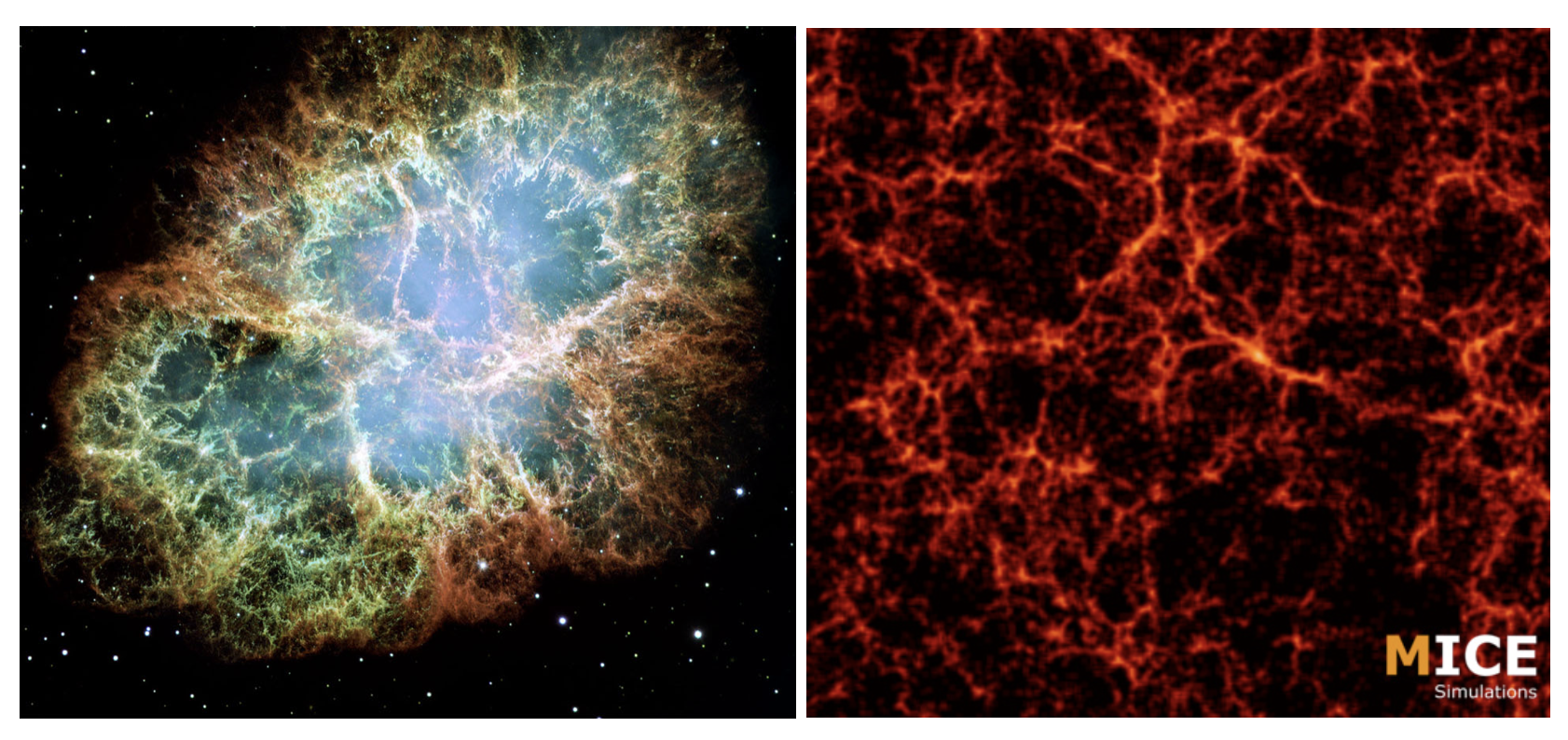}
\caption{{LEFT}: The Crab Nebula explosion as observed in 1999 from the Hubble Space Telescope, 945~years after it exploded. 
A pulsar remnant could be part of the Dark Matter.
RIGHT: the MICE simulation~\cite{Carretero2015} of our expanding Universe. The resulting structures are related in the Big Bounce model.} 
\label{fig:CrabMICE}
\end{figure}

Stars explode as supernovas (SN) either because of runaway nuclear reactions or  because of gravitational core-collapse. Protons and neutrons combine and form neutrinos by electron capture. The gravitational potential energy $\Phi$ of the collapse is converted into a neutrino burst. Neutrinos are reabsorbed by the infalling layers producing an SN explosion.
For example, the Crab Nebula pulsar in Figure~\ref{fig:CrabMICE} is thought to be a core collapse supernova that exploded releasing a total energy of $10^{51}-10^{52}$ ergs in the explosion. This energy is very similar to the FLRW collapsed energy of a $M_\odot$ star within $r_H \simeq 30$Km. Recall that the collapse speed is $c$ for $r_H$, so this is also closed to the internal (or rest) energy in Einstein's most famous equation: $E=M_\odot c^2$. 

The bounce is synchronized at different locations because the background density is the same everywhere in the FLRW cloud. The collapse  energy is converted into expansion energy ($H>0$). Neutron stars, small primordial BHs 
(PBHs) or quark stars~\cite{1970PThPh..44..291I} could result as remnants of the SN explosions and they could contribute to the Dark Matter that we see today~\cite{2020ARNPS..70..355C}. The bounce happens at times and energy densities  which are many orders of magnitudes away from Inflation or Planck times ($\tau \simeq 10^{-35}$ s or $10^{19}$ GeV). So Quantum Gravity is not needed to understand cosmic expansion and there is no monopole problem~\cite{Guth1981}.
This idea needs to be worked out and simulated.
Cold nuclear matter at neutron density is a major unsolved problem in modern physics~\cite{2016ARA&A..54..401O},
but a Big Bang from a Big Bounce seems more plausible that one that comes out of nothing. 

\subsection{The Horizon Problem}

\begin{figure}[H]
\includegraphics[width=1.\linewidth]{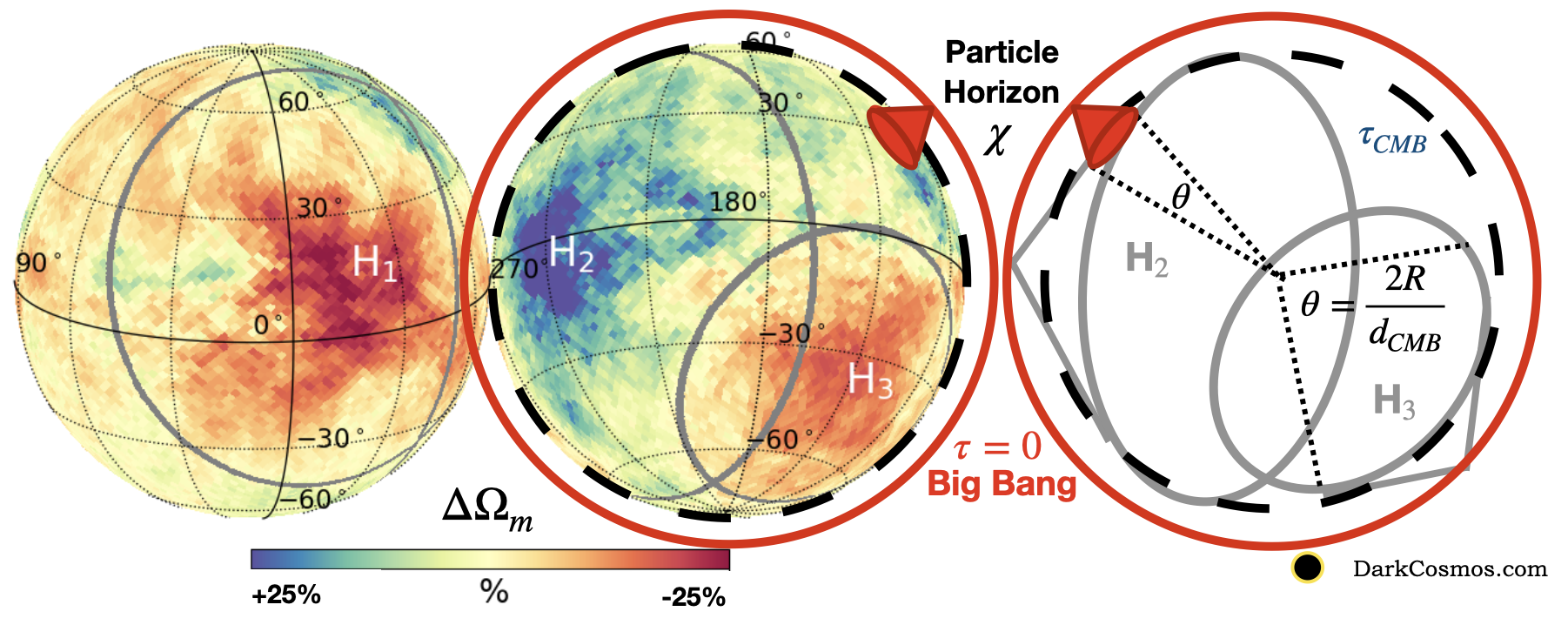}
\caption{{The CMB} sky represented as the surface of a  sphere (two view angles) whose radius is the distance traveled by the CMB light to reach us (at the center of the sphere). The red circle represents the corresponding spherical surface from the Big Bang light ($\tau=0$), if we could see it. The CMB particle horizon $\chi \sim r_H$  (small red cones) is the distance travel by light between $\tau=0$ and $\tau_{CMB}$ and subtends a very small angle in the observed CMB sky. Large grey circles on the CMB surface are therefore super-horizon boundaries (labeled $\rm{H_1}$, $\rm{H_2}$ and $\rm{H_3}$) in the relative variations of cosmological parameters (color scale) at different locations of the CMB sky~\cite{FG20}.
Regions $H_i$ correspond to a cutoff in super horizon perturbations (of size $\theta \simeq 2R/d_{CMB} \simeq 60$deg.) out of the $\tau=0$ surface. They are inside our BHU, but not causally connected (yellow region in  Figure~\ref{fig:BHevolution}).}
\label{fig:CMBhorizons}
\end{figure}

The farther back we observe an image in the sky, the older it is.
The Big Bang, if we could see it, corresponds to a very distant spherical shell in the sky, represented by large red circles in Figure~\ref{fig:CMBhorizons}. The furthest we can actually see is the CMB shell (dashed circle), which is quite close to $\tau=0$. This means that $r_H$ (or corresponding comoving particle horizon $\chi$) subtends a very small angle in the sky. So no physical mechanism can create the uniform CMB temperature that we see across the full sky. The initial conditions in the Big Bang have to be uniform to start with. This is very unlikely if the Big Bang came out of nothing~\cite{Guth1981,Dyson}. However, this is exactly what we expect if the Big Bang originates from a uniform Big Bounce. This provides a solution to the Horizon problem without Inflation.

After the Big Bang, the resulting radiation and plasma fluids cool down following the standard FLRW evolution (nucleosynthesis and CMB recombination). The Big Bounce has super horizon irregularities from the collapsing phase which re-enter the horizon $r_H$  as the expansion proceeds. These are the seeds for new structures  (BAO and galaxies) that grow under gravitational instability, as illustrated in Figure~\ref{fig:CrabMICE}.
The key difference with Inflation is that in the BHU the spectrum of incoming fluctuations have a cutoff for scales larger than $\lambda>2R$ ($k<\pi/R$), while Inflation is scale invariant in all scales. This results in an anomalous lack of the largest structures in the CMB sky temperature $T$ with respect to the predictions of Inflation. This particular CMB anomaly is well known but is often interpreted in different ways~\cite{COBEw2,WMAP,Gaztanaga2003,Efstathiou2010,Schwarz2016}.

\begin{figure}[H]
\includegraphics[width=1.\linewidth]{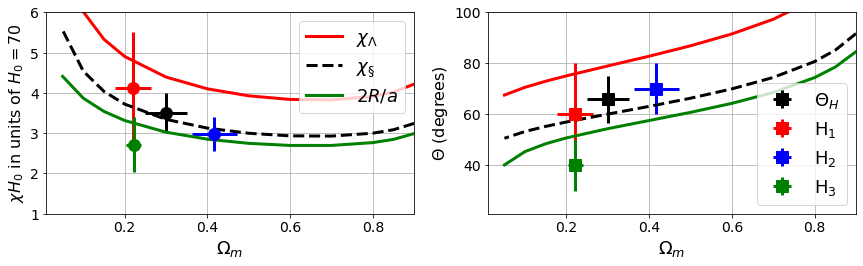}
\caption{{Comparison} of the causal horizon $H_i$ sizes shown in Figure~\ref{fig:CMBhorizons} and $\theta_H$ from the homogeneity index~\cite{Benjamin} in comoving $\chi H_0$ and angular units, given the mean  measured $\Omega_m$ and $H_0$ in each horizon. This is compared to the BHU predictions ($2R/a$, green), $\chiC$~\cite{Gaztanaga2021} (dashed) and $\chi_\Lambda =r_S/a$ (red) as a function of $\Omega_m$.}
\label{fig:CMBhorizons2}
\end{figure}

A related anomaly is shown in Figure~\ref{fig:CMBhorizons}. It displays a sky map of relative variations in the fitted values of $\rho_m$ (or cosmological parameter $\Omega_m=\Omega_B+\Omega_{DM}=1-\Omega_\Lambda$) over large regions around each position in the sky.
There is a characteristic cutoff scale (or causal horizon) shown by grey circles labeled $H_i$. Same horizons are found for different cosmological parameters.
This can be interpreted as a detection of super horizon fluctuations from the Big Bounce, with a cutoff given by the size of $H_i$. 
 Similar results were found later by an independent analysis~\cite{2022arXiv220103799Y}.
In Figure~\ref{fig:CMBhorizons2} we 
compare the different cutoff scales with the predictions of the BHU. There is a good agreement for both comoving scales  $\chi H_0$ (left panel) and angular scales.
These are independent because only the former depends on the measured values of $H_0$ in each horizon. A recent study of the homogeneity index in the CMB~\cite{Benjamin} finds a cutoff scale $\Theta_H = 66 \pm 9$ degrees. This is shown as the black symbol in Figure~\ref{fig:CMBhorizons2} for the mean values of $\Omega_m=0.3$ and $H_0=67$ Km/s/Mpc in the full CMB sky. 

\subsection{Dark Energy}

During the expanding phase we need to include $\Omega_\Lambda$ in the dynamics because the BH Event Horizon $r_S$  forbids anything to escape.  This appears in the action of GR as a Gibbons-Hawking-York (GHY) boundary term ~\cite{York,Gibbons,Hawking1996}, which is equivalent to
a $\Lambda$ term when the evolution happens inside an expanding BH (see Appendix \ref{app:lambda}). The measured $\Lambda$ term is the Event Horizon of our BHU in Equation~(\ref{eq:rs}). In the standard Big Bang model there is no reason for cosmic acceleration. A new exotic ingredient, Dark Energy (DE), has to be added to account for this new evidence. Moreover, 
there is no fundamental understanding as to why the DE equation of state $\omega \equiv \rho/p$ should be so close to $\omega = -1$ as found by the latest data compilations~\cite{DES2021}. This is instead  the natural  outcome of the BHU because  $\omega = -1$ corresponds to a constant BH Event Horizon $r_S=r_\Lambda$.
The expansion freezes and becomes static in the physical frame. For a comoving observer this looks like exponential inflation, but both pictures are equivalent~\cite{Mitra2012,hal-03344159}. As in the standard cosmological  model, it takes $14$ Gyrs to reach now (or $H_0$) from the Big Bounce. So a total of 25 Gyrs from the BH formation in Equation~(\ref{eq:tauBH}).

\section{Discusion and Conclusions}
\label{sec:conclusion}


Inflation is believed to solve the flatness problem: why our universe has a flat global topology (or geometry) $k=0$?
However, given some matter content, GR can not give us its topology. This is a global property of spacetime that is either assumed or directly measured. The same  applies to an intrinsic $\Lambda$ term. 
Equations~(\ref{eq:H2}) and (\ref{eq:R}) in the BHU are also exact solutions in GR for $k\ne \Lambda \ne 0$ by just replacing $r_H^{-2} \equiv H^2 + k/a^2 + \Lambda/3$.  However, we use $k=\Lambda=0$ here because these are the values in empty space (for Minkowski metric) and there is no reason,  within GR, that they should be different in the presence of matter.
So we believe there is no flatness or $\Lambda$ problem that needs to be solved. Such problems only arise when you include other considerations outside GR, for example when you try to explanin that the Big Bang could have emerge out of nothing or from Quantum Gravity. Which is not needed in the BHU.

The \LCDM model interprets cosmic acceleration as evidence for Dark Energy (or an intrinsic $\Lambda$ term). We have shown instead that this is an indication that  we live inside a BH of mass given by Equation~(\ref{eq:rs}).  Such interpretation provides a fundamental explanation to the meaning of $\Lambda$ (see also Appendix \ref{app:lambda}). 
This results in the BHU solution to GR (see Appendix \ref{app:exactGR}), which we have reproduced here in Equation~(\ref{eq:R}) just using simple Newtonian physics.
The idea that the universe might be generated from the inside of a BH is not new and has extensive literature~\cite{Gonzalez-Diaz,Easson,Daghigh,Firouzjahi,Yokoyama,Dymnikova} which mostly focused in deSitter  metric for the BH interior. The formation mechanisms involve some modifications or extensions of GR, often motivated by Quantum Gravity or String Theory. 
The BHU solution is also similar to the Bubble Universe and gravastar solutions~\cite{1987PhRvD..35.1747B,1989PhLB..216..272F,Aguirre,gravastar2015,Garriga16,PBH3}. However, there are no surface terms (or Bubble) in the BHU and there is regular matter and radiation inside (see Appendix \ref{app:exactGR}). Several authors before have proposed that the FLRW metric could be the interior of a BH~\cite{Pathria,Good,Poplawski,Zhang,1997lico.book.....S}. However, these solutions were incompleted~\cite{Knutsen} or outside GR. Stuckey~\cite{Stuckey} showed that a dust filled FLRW metric can be joined to an outside BH metric. This is an independent precursor to the BHU model. Note that the BHU is located within a larger spacetime and is only homogeneous inside the event horizon. We do not know much about the larger spacetime, but it is in principle possible to observe light and matter that comes from outside.
 
How did our Universe end up inside a BH? If it collapsed to form one, why is it expanding now? As illustrated in Figure~\ref{fig:BHevolution}, the BHU has a mathematical singularity at $\tau=0$. As we approach that singularity, causal regions become small so the physics involved is similar to that of stars.
In nature we have never observed stable cold matter with densities larger than that of an atomic nuclei.
We propose here that when the collapsed reaches nuclear saturation density it stops, explodes and bounces back, like a supernova. This is due to the
same  neutron degeneracy pressure that occurs in NS and atomic nuclei.
The time before $\tau=0$ represents a causal horizon which divides the BHU into smaller solar mass regions that explode and bounce in sync as long as the density is the same. Super-horizon perturbations, produced during the collapse, will be sync out of phase and generate some irregularities.
Further work is needed to understand the details and conditions for such Big Bounce to happen and to estimate the perturbations and composition and fraction of compact and difuse renmants that resulted from the SN explosions. 

This provides a uniform start for the Big Bang, solving the horizon problem. The yellow shaded regions in  Figure~\ref{fig:BHgrowth}-\ref{fig:BHevolution} show that the mass that collapsed into our BHU moved outside the horizon $r_H$.
super-horizon perturbations could seed structure (BAO and galaxies) as they re-enter $r_H$ during expansion. The main differences with Inflation is the origin of those perturbations and the existence of a cutoff in the spectrum of fluctuations given by $R(\tau)$. As illustrated in Figure~\ref{fig:CMBhorizons}-\ref{fig:CMBhorizons2}, such cutoff has  recently been measured in the CMB maps~\cite{FG20,Benjamin}. Current and future galaxy surveys are also able to measure this signal~\cite{GB1998,Barriga2001} which 
could also appear as a dipole~\cite{2021ApJ...908L..51S}. The existence of super horizon  perturbations could also be related to the tension in measurements of cosmological parameters from different cosmic scaletimes~\cite{Riess19,DiValentino,2019Natur.571..458C} which have similar variations in cosmological parameters~\cite{GCQ2022}.
The Big Bounce could also help us understand two remaining mysteries in the \LCDM paradigm: the origin for the amplitude $\delta_T \simeq 10^{-5}$ in the CMB and the nature of Dark Matter. Structure formation during the collapse and bounce could be key to understand them.  Compact remnants such as BHs or neutron could also be detected and account for all Dark Matter.
In Appendix \ref{app:rot} we give some simple considerations on the role of BH rotation. These ideas requires further work and validation. Table \ref{tab:comparison} presents a comparison of \LCDM and BHU solutions. 

\begin{table}[H]
	\centering
	\caption{Model comparison. Observations that require explanation.} 
	\label{tab:comparison}
	\setlength{\cellWidtha}{\fulllength/3-2\tabcolsep-0in}
\setlength{\cellWidthb}{\fulllength/3-2\tabcolsep-0in}
\setlength{\cellWidthc}{\fulllength/3-2\tabcolsep-0in}
	\begin{adjustwidth}{-\extralength}{0cm}
		\begin{tabularx}{\fulllength}{>{\centering\arraybackslash}m{\cellWidtha}>{\centering\arraybackslash}m{\cellWidthb}>{\centering\arraybackslash}m{\cellWidthc}}
			\toprule
\textbf{Cosmic Observation} & \textbf{Big Bang (\LCDM) Explanation} & \textbf{BHU Explanation}\\ \midrule
Expansion law & FLRW metric & FLRW metric \\
Element abundance &   Nucleosynthesis  & Nucleosynthesis \\
Cosmic Microwave Background (CMB) &  recombination  &  recombination  \\
All sky CMB uniformity  & Inflation & Uniform Big Bounce \\
Cosmic acceleration, BAO \& ISW  & Dark Energy  & BH event horizon size\\
14Gyr age since $\tau=0$ & Dark Energy  & BH event horizon size \\
Rotational curves  \& Cosmic flows  & Dark Matter  &  compact remnants (BHs, NS) of Big Bounce\\
$\Omega_m > \Omega_B$  \& gravitational lensing & Dark Matter & compact remnants (BHs, NS) of Big Bounce \\
CMB fluctuations $\delta T= 10^{-5}$ &  free parameter & Big Crunch perturbations \\ 
$\Omega_m/\Omega_B \simeq 4$ & free parameter & fraction of compact to difuse renmants \\
$\Omega_\Lambda/\Omega_m \simeq 3$ & free parameter & time to deSitter phase \\
Large scales anomalies in CMB & Cosmic Variance (bad luck) & super-horizon cutoff $\lambda <2R$ \\
anomalies in cosmological parameters & Systematic effects &  super-horizon perturbations \\
flat universe $k=0$  & Inflation & topology of empty space \\
monopole problem  & Inflation &  low energy Big Bounce \\\bottomrule
		\end{tabularx}
	\end{adjustwidth}
\end{table}

The BHU solution can also be used to understand the interior of a BH. This sounds similar to Smolin~\cite{1997lico.book.....S}, who speculated that all final (e.g. BH) singularities 'bounce' or tunnel to initial singularities of new universes. However, the bounce proposed here, based on Pauli exclusion principle in Quantum Mechanics, could avoid both the BH and the Big Bang mathematical singularity theorems~\cite{Penrose,Dadhich}. That a non singular version of such solutions exist is clear from direct observation of stars and BHs. As stated by Ellis~\cite{Ellis}, the concept of physical infinity is not a scientific one.

The Big Bounce proposed here, based in Quantum Mechanics, could avoid both the BH and the Big Bang singularities~\cite{Penrose,Dadhich}. 
The BHU also eludes the entropy paradox~\cite{Dyson} in a similar way as that proposed by Penrose~\cite{PenroseEntropy}. The difference is that the BHU does not require new laws (infinite conformal re-scaling) or cyclic repetition. 
 Our Universe will end up trapped, static and frozen, just as first modeled by Einstein in 1917~\cite{Einstein1917} when he introduced $\Lambda$. However, we have found here that $\Lambda$ is just the Event Horizon of our BHU and therefore of a larger and older background which could contain many other BHUs

\vspace{6pt}
\funding{{This work was partially supported by grants from Spain PGC2018-102021-B-100 and Unidad de Excelencia María de Maeztu CEX2020-001058-M and from European Union LACEGAL 734374 and EWC 776247. IEEC is funded by Generalitat de Catalunya.}}
\dataavailability{{No new data is presented.}}
\acknowledgments{To Benjamin Camacho-Quevedo, Pablo Fosalba, Nanda Rea, Facundo Rodriguez and Santi Serrano for comments on the original manuscript.}
\conflictsofinterest{The author declares no conflict of interest.}



\appendixtitles{yes} 
\appendixstart
\appendix

\section{Exact solution in General Relativity}
\label{app:exactGR}

The flat FLRW metric in comoving coordinates $\xi^\alpha=(\tau,\chi,\theta,\delta)$, corresponds to an homogeneous and isotropic space:
\beq
ds^2 =f_{\alpha\beta} d\xi^\alpha d\xi^\beta = -d\tau^2 + a(\tau)^2\left[ d\chi^2 + \chi^2 \dA \right],
\label{eq:frw}
\eeq
where we have introduced the solid angle: 
$\dA^2 = d\theta^2 + \sin{\theta}^2 d\delta^2$.
The scale factor, $a(\tau)$, describes the expansion/contraction as a function of comoving or cosmic  time $\tau$ 
(proper time for a comoving observer). For a perfect fluid  Equation~(\ref{eq:Tmunu})  with density $\rho$ and pressure $p$, the solution to Einstein's field equations Equation~(\ref{eq:rmunu}) is well known~\cite{Padmanabhan}:
\bea
&&  
 H^2 \equiv \left(\frac{\dot{a}}{a}\right)^2  =  \frac{8\pi G}{3} \rho =  H_0^2  \left[ \Omega_m a^{-3} + \Omega_R a^{-4} + \Omega_\Lambda \right] 
\label{eq:Hubble}
 \\ &&  
 \rho_\Lambda  \equiv \rho_{\textrm{vac}} + \frac{\Lambda}{8\pi G} 
 \label{eq:rhoL}
\,\,\,\, ; \,\,\,\,  \rho_c \equiv \frac{3H^2}{8\pi G}
\,\,\,\, ; \,\,\,\, \Omega_{X} \equiv \frac{\rho_X}{\rho_c(a=1)}, 
\eea
where $\Omega_m$ (or $\rho_m$) represent the matter density 
today ($a=1$), $\Omega_R$ is the radiation, $\rho_{\rm vac}$  represents vacuum energy: $ \rho_{\rm vac} =  -p_{\textrm{vac}}=V(\varphi)$ 
and $\rho_\Lambda=-p_\Lambda$ is the effective cosmological constant density. Note that $\Lambda$ (the raw value) 
is always constant, but $\rho_\Lambda$ (effective value) can change if $\rho_{vac}$ changes.
Given $\rho(\tau)$ and $p(\tau)$ we can use the above equations to find $a(\tau)$. 

Consider next the most general form of a
metric with spherical symmetry in physical or Schwarzschild (SW) coordinates $(t,r,\theta, \varphi)$~\cite{Padmanabhan,Dodelson}:
\beq
 ds^2 = g_{\mu\nu} dx^\mu dx^\nu =
 -(1+2\Psi) dt^2 +  \frac{dr^2}{1+2\Phi} + r^2 \dA^2,
\label{eq:newFRW}
 \eeq 
 where  $\Phi=\Phi(t,r)$ and $\Psi=\Phi(t,r)$ are the generic gravitational potentials.
 The Weyl potential $\Phi_W$ is the geometric mean of the two:
 \beq
 ( 1+2\Phi_W)^2= (1+2\Phi) (1+2\Psi).
 \eeq
$\Psi$ describes propagation of non-relativist particles and $\Phi_W$ the propagation of light. The solution to Einstein's field equations Equation~(\ref{eq:rmunu}) 
for empty space ($\rho=p=\rho_\Lambda=0$) results in the Schwarzschild (SW) metric:
 \beq
2\Phi= 2\Psi= - 2GM/r \equiv -  r_{S}/r,
\label{eq:Schwarzschild}
 \eeq 
 which describes a singular BH  of mass $M$ at $r=0$.
 The solution for $\rho=p=M=0$, but $\rho_\Lambda \ne 0$ results in deSitter (dS) metric:
  \beq
2\Phi= 2\Psi = - r^2/r_\Lambda^2 \equiv 
- r^2 H_\Lambda^2  = - r^2 8\pi G\rho_\Lambda/3.
\label{eq:deSitter}
\eeq
{We} also consider a generalization of the dS metric, which we call a dS extension (dSE), which is just a recast of the general case:
\beq
2\Phi(t,r) \equiv  -  r^2  H^2(t,r) \equiv - r^2/ r_H^2,
\label{eq:dSE}
\eeq
and arbitrary $\Psi$.

\subsection{Dual Frame: FLRW in the Physical Frame}
\label{sec:frame}

\begin{figure}[H]
\includegraphics[width=.75\linewidth]{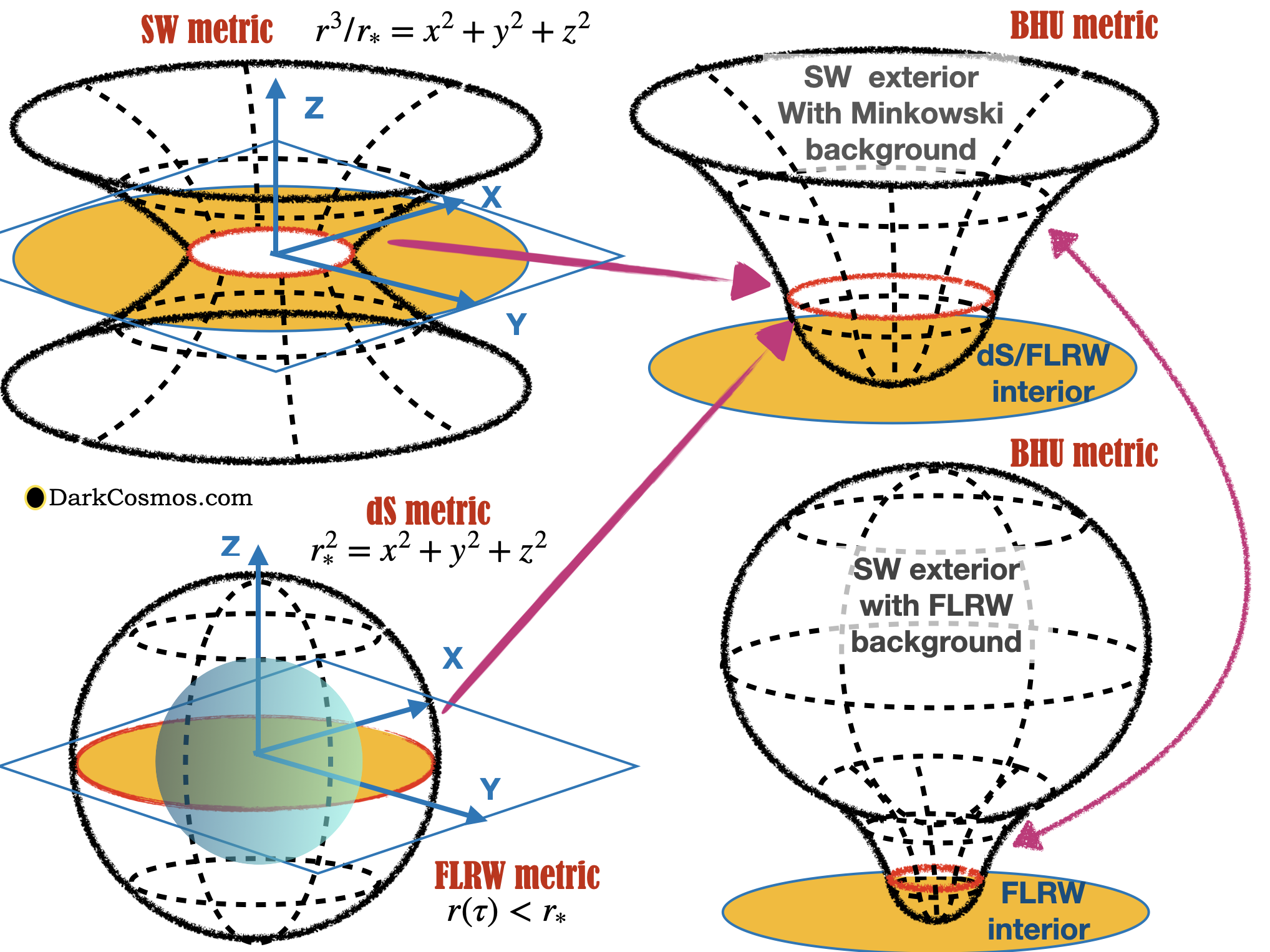}
\caption{Representation 
of $ds^2= (1+2\Phi)^{-1} dr^2 + r^2 d\theta^2$  in polar coordinates embedded in 3D flat space.
Yellow region shows the 2D projection coverage in the true $(x,y)$ plane. In bottom left  we show: deSitter (dS,  $2\Phi=-r^2/r_*^2$) and  FLRW  ($r=r(\tau)<r_*$, blue sphere inside dS). In top left, we show Schwarzschild (SW,  $2\Phi=-r_*/r$).  In the top right, we show a BHU with dS (or FLRW) interior and SW exterior.
More generally, the BHU solution has two nested FLRW metrics join by SW metric (bottom right).}
\label{fig:BHU}
\end{figure}

Consider a change of variables from  $x^\mu= [t, r]$ to comoving coordinates $\xi^\nu=[\tau,\chi]$, where $r=a(\tau) \chi$ and angular variables $(\theta, \delta)$ remain the same.
The metric  $g_{\mu\nu}$ in \mbox{Equation~(\ref{eq:newFRW})} transforms to $f_{\alpha\beta}=\Lambda^\mu_\alpha \Lambda^\nu_\beta g_{\mu\nu}$, with $\Lambda^\mu_\nu \equiv
\frac{\partial x^\mu}{\partial \xi^\nu}$. If we use:
\beq
\Lambda= \begin{pmatrix}
  \partial_\tau t &  \partial_\chi t \\
  \partial_\tau r    &  \partial_\chi r  \\
 \end{pmatrix}
=
\begin{pmatrix} 
(1+2\Phi_W)^{-1} & a r H (1+2\Phi_W)^{-1}
 \\ 
r H  &  a,
\end{pmatrix}
\label{eq:xi2xH}
\eeq
with $2\Phi= -r^2 H^2$ and arbitrary $a(\tau)$ and $\Psi$, we find: 
\beq
f_{\alpha\beta}= \Lambda^T
\begin{pmatrix} 
-(1+2\Psi)
& 0 
 \\ 
0 & (1+2\Phi)^{-1} 
\end{pmatrix}
\Lambda 
=
\begin{pmatrix} 
-1 & 0 
 \\ 
0 & a^2. 
\end{pmatrix}
\label{eq:fab}
\eeq
{In} other words,  these two metrics are the same:
 \beq
 -(1+2\Psi) dt^2 +  \frac{dr^2}{1- r^2 H^2} 
 = -d\tau^2 + a^2 d\chi^2. 
\label{eq:newFR3}
 \eeq 
The dSE metric in Equation~(\ref{eq:dSE}) with $2\Phi= -r^2 H^2$ corresponds to the FLRW metric with $H(t,r)= H(\tau)$: this is a hypersphere of radius $r_H$ that tends to $r_\Lambda$ (see bottom left panel in Figure~\ref{fig:BHU}). 
This frame duality can be understood as a Lorentz contraction $\gamma=1/\sqrt{1-u^2}$ where the velocity $u=\dot{r}$ is given by the expansion law: $\dot{r}= Hr$. An observer in the SW frame, not moving with the fluid, sees the moving fluid element $a d\chi$ contracted by the Lorentz factor $\gamma$:  $a d\chi \Rightarrow \gamma dr $. 

\subsection{The BHU Solution}

We next look for solutions where we have matter $\rho_m=\rho_m(t,r)$ and radiation $\rho_R=\rho_R(t,r)$ inside some radius $R$ and empty space outside:
\bea
\rho(t,r) &=& \left\{ \begin{array}{ll} 
0  &  {\text{for}} ~ ~ r>R \\
\rho_m+ \rho_R& {\text{for}} ~ ~r<R \\
\end{array} \right. .
\label{eq:rhoVu}
\eea
 {When} $R>r_S$, we call this an FLRW cloud and when $R<r_S$ this is a BH Universe. For $r>R$, we have the SW metric. 
For the interior we use the dSE notation in Equation (\ref{eq:dSE}): $2\Phi(t,r) \equiv -r^2 H^2(t,r) \equiv -r^2/r_H^2$, so that:
\beq
2\Phi(t,r) = \left\{ \begin{array}{ll} 
  - r_{S}/r  &  {\text{for}} ~ ~ r>R\\
 -r^2 H^2  & {\text{for}} ~ ~r<R \\
\end{array} \right. .
\label{eq:BH.u2}
\eeq
{At} the junction $r=R$, we reproduce Equation~(\ref{eq:R}).
For $r<R$, we can change variables as in Equations~(\ref{eq:xi2xH})--\ref{eq:newFR3}). 
In the comoving frame of Equation~(\ref{eq:newFR3}), from every point inside de BHU, comoving observers will have the illusion of an homogeneous and isotropic space-time around them, with a fixed Hubble--Lemaitre expansion $H(\tau)$. This converts dSE metric into FLRW metric. 
So the solution is $H(t,r)=H(\tau)$
and $R(\tau)=[r_S/H^2(\tau)]^{1/3}$.
Given $\rho(\tau)$ and $p(\tau)$ in the interior we can use Equation~(\ref{eq:Hubble})
to find $H(\tau)$ and $R(\tau)$:
\beq
H^2(\tau) = \frac{8\pi G}{3} \rho(\tau) = \frac{r_S}{ R^3(\tau)}.
\eeq
{This} corresponds to a homogeneous FLRW cloud of fix mass $M=r_S/2G$ 
in Equations~(\ref{eq:H2}) and (\ref{eq:R}). This solution exist for any content inside $R$
~\cite{hal-03344159}.  Figure~\ref{fig:BHU} shows a spatial representation of the SW, dS, FLRW and BHU solutions.

\subsection{Junction Conditions}

We can arrive at the same BHU solution using Israel's  junction conditions (~\cite{Israel,Barrabes1991}). We can combine two solutions  with different energy content,  as in Equation~(\ref{eq:rhoVu}), to find a new solution. To do that we need to find 
a hypersurface junction $\Sigma$ to match them well. In our case, this will be given by $R$. The junction conditions require that the metric and its derivative (the extrinsic curvature $K$) match at  $\Sigma$. The join metric then provides a new solution to GR. In many cases, like in the Bubble Universes or gravastar ~\cite{1987PhRvD..35.1747B,1989PhLB..216..272F,Aguirre,gravastar2015,Garriga16,PBH3}, which match dS and SW metric, this does not work and the junction requires a surface term (the bubble) to glue both solutions together.
For the BHU there are no surface terms~\cite{hal-03344159}, which shows that this is an exact solution. In the limit where the FLRW has constant $H$ (i.e., our future), the BHU solution corresponds to match between dS and SW metric. So a Bubble Universe without bubble.

To see this, consider the case where $\Sigma$ is given by $R$ in the freefall collapse of an FLRW cloud of fixed mass $M$. For matter domination, this corresponds to $R= a(\tau) \chiSW$ as in Equation~(\ref{eq:R}), where $\chiSW=r_S/a_{BH}$ is fixed.
The induced 3D metric on $\Sigma$ is $h_{\alpha\beta}^\in$ with coordinates $dy^\alpha=(d\tau,d\delta,d\theta)$:
\beq
ds^2_{\Sigma^\in}= h_{\alpha\beta}^\in  dy^\alpha dy^\beta= -d\tau^2 + a^2(\tau) \chiSW ^2 \dA^2.
\label{eq:Sigma}
\eeq
{For} the outside SW frame, the junction $\Sigma^{\out}$ is described by $r=R(\tau)$ and $t=T(\tau)$, where $\tau$ is the FLRW comoving time and $t$ the time in the physical frame. We then have:
\beq
dr= \dot{R} d\tau ~~;~~ dt= \dot{T} d\tau,
\label{eq:RT}
\eeq
where the dot refers to derivatives with respect to $\tau$.  The metric $h^{\out}$ induced in the outside SW metric is:
\bea
ds^2_{\Sigma^\out} &=& h_{\alpha\beta}^\out  dy^\alpha dy^\beta =
-F dt^2 + \frac{dr^2}{F} + r^2  \dA^2   \nonumber \\
&=& - (F\dot{T}^2- \dot{R}^2/F) d\tau^2 + R^2 \dA^2,  \label{eq:dS+}
\eea
where $F \equiv 1- r_{S}/R$. Comparing Equation~(\ref{eq:Sigma})
with Equation~(\ref{eq:dS+}), the first matching
conditions $h^{\in}=h^{\out}$ are then:
\beq
R(\tau) = a(\tau) \chiSW ~~;~~   F\dot{T} =  \sqrt{\dot{R}^2+F} \equiv \beta({R,\dot{R}}).
\label{eq:matching}
\eeq
{For} any given $a(\tau)$ and $\chiSW$ we can find both $R(\tau)$ and $\beta(\tau)$. 
We also want the derivative of the metric to be continuous at $\Sigma$. For this, we  estimate the extrinsic curvature $K^\pm$ normal to $\Sigma^\pm$ from each side of the hypersurface:
\beq
K_{\alpha\beta} = 
-\left[ \partial_a n_b - n_c \Gamma^c_{ab} \right]  e^a_\alpha e^b_\beta, 
\label{eq:Kab}
\eeq
where $e^a_\alpha =\partial x^a/\partial y^\alpha$ and $n_a$ is the 4D vector normal to $\Sigma$.
The outward 4D velocity is $u^a = e_\tau^a = (1,0,0,0)$ and
the normal to $\Sigma^\in$ on the inside is then 
$n^\in= (0,  a,0,0)$.
On the outside  $u^a = (\dot{T},\dot{R},0,0)$ and
$n^\out =(-\dot{R},\dot{T},0,0)$. It is straightforward to verify that: $n_a u^a =0$ and $n_a n^a =+1$ (for a timelike surface) for both $n^\in$ and $n^\out$. 
We then find that the extrinsic curvature in Equation~(\ref{eq:Kab}) to the $\Sigma$ junction, estimated with the inside FLRW metric, i.e., $K^\in$ is:
\bea
K^\in_{\tau \tau} &=&  -(\partial_\tau n^\in_\tau -  a \Gamma_{\tau\tau}^{\chi}) e_\tau^\tau e_\tau^\tau =0    \nonumber \\
K^\in_{\theta \theta} &=&     a \Gamma_{\theta\theta}^{\chi}  e_\theta^\theta e_\theta^\theta= -a \chi_* = -R. 
\label{eq:Kin}
\eea
{For} the SW metric:
\bea
K^\out_{\tau\tau}   &=&  \ddot{R} \dot{T} - \dot{R} \ddot{T} +\frac{\dot{T} r_{S}}{2 R^2 F} (\dot{T}^2 F^2-3 \dot{R}^2) =
\frac{\dot{\beta}}{\dot{R}}  \nonumber  \\ 
 K^\out_{\theta\theta} &=&   \dot{T} \Gamma_{\theta\theta}^r =  -\dot{T} F R =  -\beta R, 
\label{eq:Kout}
\eea
where we have used the definition of $\beta$ in Equation~(\ref{eq:matching}). In both cases
$K_{\delta\delta} = \sin^2{\theta} K_{\theta\theta}$, so that when $K_{\theta\theta}^\in = K_{\theta\theta}^\out$,
it follows that  $K_{\delta\delta}^\in = K_{\delta\delta}^\out$.
Comparing Equation~(\ref{eq:Kin}) with Equation~(\ref{eq:Kout}), the  matching conditions $K^{\in}_{\alpha\beta}=K^{\out}_{\alpha\beta}$ require $\beta=1$, which using Equation~(\ref{eq:matching}) gives:$ R= \left[r_H^2 r_{S} \right]^{1/3}$.
This reproduces the junction in Equation~(\ref{eq:R}). So the two metrics and  derivatives (the extrinsic curvature) are identical in the hypersurface defined by $R$. This completes the proof that the FLRW cloud is an exact solution of GR without surface terms. For more details see~\cite{hal-03344159}.

\section{The Action of GR and the $\Lambda$ term}
\label{app:lambda}

Consider the Einstein--Hilbert action (~\cite{Hilbert1915,Padmanabhan}):
\beq
S = \int_\calMa d\calMa \left[ \frac{ R-2\Lambda}{16\pi G} +  {\cal L} \right],
\label{eq:action}
\eeq
where $d\calMa=\sqrt{-g} d^4 x$ is the invariant volume element, $\calMa$ is the volume of the 4D spacetime manifold, $R= R^\mu_\mu = g^{\mu\nu} R_{\mu\nu}$ is the Ricci scalar curvature and ${\cal L}$ the Lagrangian of the energy-matter content. We can obtain Einstein's field equations (EFE) for the metric field $g_{\mu\nu}$  from this action by requiring $S$ to be stationary $\delta S=0$ under arbitrary variations of the metric $\delta g^{\mu\nu}$. The solution  is (~\cite{Einstein1916,Padmanabhan}):

\beq
 G_{\mu\nu}+\Lambda g_{\mu\nu}=
 8\pi G~T_{\mu\nu} \equiv - \frac{16\pi G}{\sqrt{-g}} \frac{\delta (\sqrt{-g} {\cal L}) }{\delta g^{\mu\nu}},
\label{eq:rmunu}
\eeq
where $G_{\mu\nu} \equiv  R_{\mu\nu} - \frac{1}{2} g_{\mu\nu} R $.
For perfect fluid in spherical coordinates:
\beq
T_{\mu\nu} =  (\rho+p) u_\mu u_\nu + p g_{\mu\nu},   
\label{eq:Tmunu}
\eeq 
where $\rho$,
and $p$ are the energy-matter density and pressure.
This fluid can be made of several components, each with a different equation of state $p=\omega \rho$.

Equation~(\ref{eq:rmunu}) requires that boundary terms vanish (e.g. see~\cite{Landau1971,Padmanabhan}). If there are boundaries to the dynamic equations, we need to add a Gibbons-Hawking-York (GHY) boundary term ~\cite{York,Gibbons,Hawking1996} to the action in Equation~(\ref{eq:action}):
\beq
S_{GHY}= \frac{1}{8\pi G} \oint_{\partial \calMa}   d^3y \sqrt{-h} \,K,
\label{eq:actionGHY}
\eeq
so that the total action is $S+S_{GHY}$ and $K$ is the trace of the extrinsic curvature at the boundary $\partial \calMa$ and $h$ is the induced metric. 
The expansion that happens inside an isolated BH is bounded by its event horizon $r<r_S$ and we need to add the GHY boundary term  $S_{GHY}$.
The integral is over the induced metric at ${\partial \calMa}$, which for a time-like junction $d\chi=0$
corresponds to $R=r_S$:
\beq
ds^2_{{\partial \calMa}}= h_{\alpha\beta}  dy^\alpha dy^\beta= -d\tau^2 + r_S^2 \dA^2.
\label{eq:SigmaR}
\eeq
{So} the only remaining degrees of freedom in the action are time $\tau$ and the angular coordinates.
We can use this metric and the trace of the extrinsic curvature at $R=r_S$
to estimate $K = -2/r_S$ from Equation~(\ref{eq:Kin}).
This result is also valid for a null geodesic~\cite{hal-03344159}. 
We then have:
\beq
S_{GHY}= \frac{1}{8\pi G} \int d\tau \, 4\pi r_S^2 \, K = - \frac{r_S}{G} \tau.
\label{eq:actionGHY2}
\eeq
{The} $\Lambda$ contribution to the action in   Equation~(\ref{eq:action}) is:
$S_{\Lambda}= - \Lambda \calMa/(8\pi G) =   - r_S^3 \Lambda \tau /3G, $
where we have estimated the total 4D volume $ \calMa $ as that bounded by  $\partial \calMa$ inside $r<r_S$. i.e.,: $ \calMa = 2V_3\tau$, where the factor 2 accounts for the fact that  $V_3= 4\pi r_S^3/3 $  is covered twice, first during collapse and again during expansion. 
Comparing the two terms we can see that we need $\Lambda =3 r_S^{-2} $ or equivalently $\rL=r_S$ to cancel the boundary term. In other words: expansion inside a BH event horizon induces an effective $\Lambda$ term in the EFE even when there is no $\Lambda$ background term to start with.
Such event horizon becomes a boundary for outgoing geodesics, i.e., expanding solutions. 
This provides a fundamental interpretation of the observed $\Lambda$ as a causal boundary~\cite{Gaztanaga2020,Gaztanaga2021,GaztaUniverse}.

\section{Outside Our BHU: A Rotating  Cloud}
\label{app:rot}

If the FLRW cloud is not totally isolated it could have some rotation. This could be a way to infer if there is something outside our BHU. Any rotation, no matter how small, could prevent or interfere with the cloud collapse. Can we detect such rotation?  A rotating BH is a bit more difficult to model because spherical symmetry is lost and the BH becomes oblate (i.e., the Kerr metric~\cite{Kerr}):
\beq
x = \sqrt{r^2 + r_J^2} \sin{\theta} \cos{\Phi} \,\,\,\,;\,\,\,\,
y = \sqrt{r^2 + r_J^2} \sin{\theta} \cos{\Phi}   \,\,\,\,;\,\,\,\,
z = r\cos{\theta}, 
\eeq
where $r_J = J/M$ is the ratio between the angular momentum $J$ and the BH mass. A detailed analysis of this case is outside the scope of this review, but we will make some energy considerations to understand the possible impact of such rotation on the Big Bounce.
 We assume that both mass $M$ and angular momentum $J$ are conserved, so $r_J$ is constant.
We also assume that $r_J \ll r_S$ so during the collapse we can neglect deviations from spherical symmetry. If we start from the FLRW cloud of size $R$ and mass $M$ with some  small initial rotation, $\dot{\theta}$, these products have to be constant:
\beq
\frac{J}{M} = r_J = R^2 \dot{\theta}  = r_S^2 \dot{\theta}_{BH}. 
\label{eq:JM}
\eeq
{As} $R$ gets smaller, $\dot{\theta}$ will become larger.
The kinetic energy term in Equation~(\ref{eq:phi})  will have another contribution $2K= \dot{R}^2+ \dot{\theta}^2 R^2$, so that Equation~(\ref{eq:H2}) becomes:
\beq
r_H^{-2} \equiv  H^2(\tau) = \frac{8\pi G}{3} \rho(\tau)  -\frac{r_J^2}{R^4} 
=  
r_S^{-2} \left(\frac{a}{a_{BH}}\right)^{-3(1+\omega)} 
-\frac{r_J^2}{r_S^4} \left(\frac{a}{a_{BH}}\right)^{-4(1+\omega)}, 
\label{eq:H2rot}
\eeq
where in the last step we have used Equations~(\ref{eq:R}) and  (\ref{eq:collapse})
for a collapsing FLRW cloud with equation of state $\omega$. So, for $\omega=0$, rotation acts like a radiation term of negative energy density. Rotation is negligible, except
when $a \Rightarrow 0$ when rotation tends to delay the collapse, as it reduces the expansion rate  $H$. Unless angular momentum is lost some other way, the rotation component will dominate (stop the collapse) for:
\beq
r_J \simeq  r_S  \left(\frac{a}{a_{BH}}\right)^{(1+\omega)/2}. 
\eeq
{Close} to the Big Bounce, if  radiation dominates ($\omega =1/3$) with  neutron  energy densities (GeV), we have $a \simeq 10^{-12} a_{BH}$. So the condition for the rotation not to interfere with the collapse is:
\beq
r_J \ll 10^{-8}  r_S.
\eeq
{Equivalently},  as $r_S \simeq H_0^{-1}$, see Equation~(\ref{eq:rs}), $\dot{\theta}_{BH}$ in Equation~(\ref{eq:JM}) has to be: 
\beq
\dot{\theta}_{BH} \ll 10^{-8} H_0, 
\eeq
so less than $10^{-8}$ cycles per Hubble time. 
Such a small contribution is undetectable in today's expansion law: $\Omega_{J} \simeq 10^{-16}$ in Equation~(\ref{eq:H2rot}), or during recombination, but it could be bounded using nucleosynthesis or by its affects on the Big Bounce.

\begin{adjustwidth}{-\extralength}{0cm}

\reftitle{References}

\end{adjustwidth}

\end{document}